\documentstyle[amssymb,aps,preprint]{revtex}

\begin{document}
\draft
\author{A.L. Cornelius}
\address{Department of Physics, University of Nevada, Las Vegas, Nevada,\\
89154-4002}
\author{P.G. Pagliuso, M.F. Hundley, and J.L. Sarrao}
\address{Materials Science and Technology Division, Los Alamos National\\
Laboratory,\\
Los Alamos, NM\ 87545}
\title{Field-induced magnetic transitions in the quasi-two-dimensional
heavy-fermion antiferromagnets Ce$_{n}$RhIn$_{3n+2}$ ($n=1$ or $2)$}
\date{\today}
\maketitle

\begin{abstract}
We have measured the field-dependent heat capacity in the tetragonal
antiferromagnets CeRhIn$_{5}$ and Ce$_{2}$RhIn$_{8}$, both of which have an
enhanced value of the electronic specific heat coefficient $\gamma \sim 400$
mJ/mol-Ce K$^{2}$ above $T_{N}$. For $T<T_{N},$ the specific heat data at
zero applied magnetic field are consistent with the existence of an
anisotropic spin-density wave opening a gap in the Fermi surface for CeRhIn$%
_{5},$ while Ce$_{2}$RhIn$_{8}$ shows behavior consistent with a simple
antiferromagnetic magnon. From these results, the magnetic structure, in a
manner similar to the crystal structure, appears more two-dimensional in
CeRhIn$_{5}$ than in Ce$_{2}$RhIn$_{8}$ where only about 12\% of the Fermi
surface remains ungapped relative to 92\% for Ce$_{2}$RhIn$_{8}$. When $%
B||c, $ both compounds behave in a manner expected for heavy fermion systems
as both $T_{N}$ and the electronic heat capacity decrease as field is
applied. When the field is applied in the tetragonal basal plane ($B||a$),
CeRhIn$_{5} $ and Ce$_{2}$RhIn$_{8}$ have very similar phase diagrams which
contain both first- and second-order field-induced magnetic transitions .
\end{abstract}


\section{Introduction}

Ce$_{n}$RhIn$_{2n+3}$ ($n=1$ or 2) crystallize in the quasi-two-dimensional
(quasi-2D) tetragonal structures Ho$_{n}$CoGa$_{2n+3}$, and both are
moderately heavy-fermion antiferromagnets ($\gamma \sim 400$ mJ/mol-Ce K$%
^{2} $ for both systems above $T_{N}=3.8$ K for$\ n=1$ and 2.8 K for $n=2$).
The evolution of the ground states of CeRhIn$_{5}$ as a function of applied
pressure, including a pressure-induced first-order superconducting
transition at 2.1 K, is unlike any previously studied heavy-fermion system
and is attributed to the quasi-2D crystal structure \cite{Hegger00}. In a
similar manner to CeRhIn$_{5}$, $T_{N}$ is seen to change only slightly with
pressure and abruptly disappear in Ce$_{2}$RhIn$_{8}$, though
superconductivity has not yet been observed \cite{Thompson01}.

A previous zero field heat capacity study on CeRhIn$_{5}$ revealed that the
anisotropic crystal structure leads to a quasi-2D electronic and magnetic
structure \cite{Cornelius00}. We have performed measurements of the heat
capacity in applied magnetic fields for CeRhIn$_{5}$ and Ce$_{2}$RhIn$_{8}$
in an attempt to further understand the electronic and magnetic properties
of these compounds. We find that for magnetic fields applied along the
tetragonal $c$-axis, both systems behave like typical heavy-fermion
compounds \cite{Stewart84} as both $T_{N}$ and $\gamma _{0}$ decrease as
field is increased. Very different behavior is seen when the field is
directed along the $a$-axis as $T_{N}$ is found to increase and numerous
field-induced transitions, both of first- and second-order, are observed.
These transitions correspond to magnetic field-induced changes in the
magnetic structure. In agreement with what one might expect, the magnetic
properties seem less 2D as the crystal structure becomes less 2D going from
single layer CeRhIn$_{5}$ to double layer Ce$_{2}$RhIn$_{8}$ (note that as $%
n\rightarrow \infty $, one gets the 3D cubic system CeIn$_{3}$).

\section{Results}

Single crystals of Ce$_{n}$RhIn$_{2n+3}$ were grown using a flux technique
described elsewhere \cite{Canfield92}. The residual resistivity ratio
between 2~K and 300~K using a standard 4-probe measurement and was found to
be greater than 100 for all measured crystals, indicative of high-quality
samples. A clear kink was observed in the resistivity at $T_{N}=3.8$ K for
CeRhIn$_{5}$ and $T_{N}=2.8$ K in Ce$_{2}$RhIn$_{8}$. Magnetization
measurements have shown that the effective magnetic moment is slightly
reduced from the value expected for Ce$^{3+}$ due to crystal fields \cite%
{Thompson01}. The specific heat was measured on a small ($\sim 10$~mg)
sample employing a standard thermal relaxation method.

The single crystals were typically rods with dimensions from 0.1 to 10 mm
with the long axis of the rod found to be along the $\langle 100\rangle $
axis of the tetragonal crystal. The samples were found to crystallize in the
primitive tetragonal Ho$_{n}$CoGa$_{2n+3}$-type structure \cite%
{Grin79,Grin86} with lattice parameters of $a=0.4652(1)$ nm and $c=0.7542(1)$
nm for $n=1$ and $a=0.4665(1)$ nm and $c=1.2244(5)$ nm for $n=2$ \cite%
{Hegger00,Thompson01}. The crystal structure of Ce$_{n}$RhIn$_{2n+3}$ can be
viewed as (CeIn$_{3}$)$_{n}$(RhIn$_{2}$) with alternating $n$ cubic (CeIn$%
_{3}$) and one (RhIn$_{2}$) layers stacked along the $c$-axis. By looking at
the crystal structure, we would expect that AF correlations will develop in
the (CeIn$_{3}$) layers in a manner similar to bulk CeIn$_{3}$ \cite%
{Lawrence80}. The\ AF\ (CeIn$_{3}$) layers will then be weakly coupled by an
interlayer exchange interaction through the (RhIn$_{2}$) layers which leads
to a quasi-2D magnetic structure. This has been shown to be true as the
moments are AF ordered within the tetragonal basal $a$-plane but display a
modulation along the $c$-axis which is incommensurate with the lattice for
CeRhIn$_{5}$ \cite{Bao00}. As $n$ is increased, the crystal structure should
become more 3D ($n=\infty $ being the 3D cubic system CeIn$_{3}$) and the
effects of the interlayer coupling should become less important causing the
magnetic and electronic structure to be more 3D. Indeed, the magnetic
structure in Ce$_{2}$RhIn$_{8}$ does not display an incommensurate spin
density wave (SDW) \cite{Bao01}.

The zero field data from specific heat measurements are shown in Fig. \ref%
{cp}. A peak at $T_{N}$ is clearly seen for both samples, indicating the
onset of magnetic order. The entropy associated with the magnetic transition
is $\symbol{126}0.3R\ln 2$ with the remaining $0.7R\ln 2$ recovered by 20 K
for both $n=1$ and 2. For $T>T_{N}$ the data could not be fit by simply
using $C/T=\gamma +\beta _{l}T^{2},$where $\gamma $ is the electronic
specific heat coefficient and $\beta _{l}$ is the lattice Debye term. As
found previously, one needs to use isostructural, nonmagnetic La$_{n}$RhIn$%
_{2n+3}$ to subtract the lattice contribution to $C$ \cite{Hegger00}. After
subtracting the lattice contribution, it is still difficult to extract a
value of $\gamma $ from the data. However, by performing a simple entropy
balance construction, a value of $\gamma \approx 400$ mJ/mol-Ce K$^{2}$ is
found for both $n=1$ and 2.

For temperatures below $T_{N}$, as found before on CeRhIn$_{5}$ \cite%
{Cornelius00}, the magnetic heat capacity data, where the corresponding La
compound is used to subtract the lattice contribution, can be fit using the
equation%
\begin{equation}
C_{m}/T=\gamma _{0}+\beta _{M}T^{2}+\beta _{M}^{\prime }\left(
e^{-E_{g}/k_{B}T}\right) T^{2}  \label{magnon}
\end{equation}%
where $\gamma _{0}$ is the zero temperature electronic term, $\beta
_{M}T^{2} $ is the standard AF magnon term, and the last term is an
activated AF\ magnon term. The need for an activated term to describe heat
capacity data been seen before in other Ce and U compounds \cite%
{Cornelius00,Bredl87,Dijk97,Murayama97}, and the term rises from an AF SDW
with a gap in the excitation spectrum due to anisotropy. As discussed
previously, the CeRhIn$_{5}$ magnetic structure indeed displays an
anisotropic SDW with modulation vector (1/2,1/3,0.297) \cite{Bao00} which is
consistent with this picture. The inset to Fig. \ref{cp} shows the data for $%
T<T_{N}$ and the lines are fits to Eq. \ref{magnon} for $%
T^{2}<(0.85T_{N})^{2}$. We find that the the activated term is NOT\
necessary to fit the Ce$_{2}$RhIn$_{8}$ data. Rather, there appears to be a
small feature in the heat capacity centered around \symbol{126}2 K whose
origin is unknown. This feature is also observed in transport measurements
and is known to persist as a function of pressure \cite{Sidorov01}. A
summary of the fit parameters, along with data on CeIn$_{3}$ \cite{Berton79}%
, is given in Table I. These results lead us to the conclusion that the
magnetically ordered state in CeRhIn$_{5}$ consists of an anisotropic SDW
that opens up a gap on the order of 8 K in the Fermi surface, while no such
gap is seen in Ce$_{2}$RhIn$_{8}$, consistent with its commensurate
structure. Note that the values for CeRhIn$_{5}$ are slightly different from
a previous report where the lattice contribution from LaRhIn$_{5}$ was not
subtracted from the raw data \cite{Cornelius00}. From the ratio of the
electronic contribution for temperatures above and below $T_{N},$ we
estimate that approximately $\gamma _{0}/\gamma \sim 0.12$ (12\%) of the
Fermi surface remains ungapped below $T_{N}$ for CeRhIn$_{5}$ while for Ce$%
_{2}$RhIn$_{8}$ 92\% of the Fermi surface remains ungapped. The results on
CeIn$_{3}$ of Berton {\it et al.} are also shown in Table I for comparison
where $\gamma _{0}/\gamma \sim 0.056$ (5.6\%). Clearly, the electronic
structure, as evidenced by the ratio $\gamma _{0}/\gamma ,$ becomes more 3D
in the Ce$_{n}$RhIn$_{2n+3}$ series as $n$ is increased \cite{Berton79}.

Fig. \ref{c115} shows the heat capacity in applied magnetic fields for $B||c$
for both compounds (CeRhIn$_{5}$ on the top and Ce$_{2}$RhIn$_{8}$ on the
bottom). As the magnetic moments are known to lie within the $a$-plane CeRhIn%
$_{5}$, the magnetic field is perpendicular to the magnetic moments in the
ordered state in this orientation. The applied field is not sufficient to
cause a field-induced magnetic transition in either compound. Rather, it is
found that $T_{N}$ and $\gamma _{0}$ decrease as $B$ is increased as is
usually observed in heavy fermion systems \cite{Stewart84}.

Fig. \ref{a115} shows the heat capacity in applied magnetic fields for $B||a$%
. For both samples, the Neel point (the onset of antiferromagnetic order),
and magnetic field-induced transitions of both first- ($T_{1}$) and
second-order ($T_{2}$) are clearly observed. The complete phase diagrams for
both CeRhIn$_{5}$ and Ce$_{2}$RhIn$_{8}$ showing the various observed
transitions are plotted in Fig. \ref{phasediag}. The open symbols correspond
to second order transitions ($T_{N}$ and $T_{2}$) and the filled symbols
represent the first order transition ($T_{1}$) \cite{firstord}. The dashed
lines are merely guides for the eyes. Remarkably, the phase diagrams for
both CeRhIn$_{5}$ and Ce$_{2}$RhIn$_{8}$ are extremely similar. Region I
corresponds to the standard modulated spin density wave that is
incommensurate with the lattice as reported previously in CeRhIn$_{5}$ \cite%
{Bao00}. The nature of the first- and second- order transitions going from
Region I to Regions II and III are not know, and work is underway to
determine the magnetic structures in Regions II and III. For Ce$_{2}$RhIn$%
_{8}$, Region II\ terminates at 70 kOe while for CeRhIn$_{5}$ it extends
beyond the highest measured field of 90 kOe. The first order transition
going from Region I\ or Region II to Region III is the same hysteretic field
induced transition observed in magnetization measurements on CeRhIn$_{5}$
where an increase in the magnetization of $\lesssim 0.006$ $\mu _{B}/$Ce %
\cite{Cornelius00}. These results clearly show that Ce$_{2}$RhIn$_{8}$ has
some 2D electronic and magnetic character.

\section{Conclusion}

In summary, we have measured the anisotropic heat capacity in applied
magnetic fields in the quasi-2D heavy-fermion antiferromagnets CeRhIn$_{5}$
and Ce$_{2}$RhIn$_{8}$. The magnetic and electronic properties of CeRhIn$%
_{5} $ can be well explained by the formation of an anisotropic SDW, leading
to a 2D electronic and magnetic structure. The phase diagram of magnetic
field-induced magnetic transitions is remarkably similar for both systems as
both a first- and second-order transition are observed in both compounds
when the magnetic field is along the tetragonal $a$-axis. From the heat
capacity measurements, we estimate that \symbol{126}12\%(92\%) of the Fermi
surface remains ungapped below the magnetic ordering temperature for CeRhIn$%
_{5}$ (Ce$_{2}$RhIn$_{8}$). The 2D nature of the electronic properties is a
result of the tetragonal crystal structure of Ce$_{n}$RhIn$_{2n+3}$ which
consists of $n$ cubic (CeIn$_{3}$) blocks which are weakly interacting along
the $c$-axis through a (RhIn$_{2}$) layer. Not surprisingly, it appears that
the case of $n=1$ (CeRhIn$_{5}$) is more anisotropic than the double layered
Ce$_{2}$RhIn$_{8}$ where the amount of ungapped Fermi surface below $T_{N}$
is greater than 90\% , as is found for the 3D system $(n=\infty )$ CeIn$_{3}$%
. The similarity of the field-induced transitions in both CeRhIn$_{5}$ and Ce%
$_{2}$RhIn$_{8}$ clearly shows that there is still some 2D character to the
magnetic properties of Ce$_{2}$RhIn$_{8}$. The results here shed light on
the unusual magnetic and electronic structure of CeRhIn$_{5}$ and Ce$_{2}$%
RhIn$_{8}$.

Work at UNLV is supported by DOE/EPSCoR Contract No. ER45835. Work at LANL
is performed under the auspices of the U.S. Department of Energy.

\begin{table}[tbp]
\caption{Calculated zero field heat capacity fit parameters for Ce$_{n}$RhIn$%
_{2n+3}$. $^*$Data for CeIn$_3$ is taken from Ref. 15. The various
parameters are defined in the text. Units for $\protect\gamma _{0}$ and $%
\protect\gamma $ are mJ/mol Ce-K$^{2}$ and for $\protect\beta _{M}$ and $%
\protect\beta_{M}^{\prime }$ are mJ/mol Ce-K$^{4}$. }
\label{table}\narrowtext                         
\begin{tabular}{llcccccccc}
&  & $n$ & $T_{N}$ (K) & $\gamma _{0}$ & $\gamma $ & $\beta _{M}$ & $\beta
_{M}^{\prime }$ & $E_{g}/k_{B}$ (K) &  \\ 
\tableline & CeRhIn$_5$ & 1 & 3.72 & 56 & 400 & 24.1 & 706 & 8.2 &  \\ 
& Ce$_2$RhIn$_8$ & 2 & 2.77 & 370 & 400 & 93.2 & - & - &  \\ 
& $^*$CeIn$_3$ & $\infty$ & 10 & 136 & 144 & 15 & - & - & 
\end{tabular}%
\end{table}

\widetext

\begin{figure}[tbp]
\caption{The zero field magnetic specific heat $C_m$ divided by temperature $%
T$ versus $T^{2}$ for CeRhIn$_5$ and Ce$_2$RhIn$_8$. The inset displays the
range $T^{2}<15$ K$^{2}$ corresponding to temperatures below $T_N$. The
lines are fits described in the text.}
\label{cp}
\end{figure}

\begin{figure}[tbp]
\caption{The magnetic specific heat $C_m$ divided by temperature $T$ versus $%
T$ for CeRhIn$_5$ (top) and Ce$_2$RhIn$_8$ (bottom) in various applied
fields applied along the $c$-axis.}
\label{c115}
\end{figure}

\begin{figure}[tbp]
\caption{The magnetic specific heat $C_m$ divided by temperature $T$ versus $%
T$ for CeRhIn$_5$ (top) and Ce$_2$RhIn$_8$ (bottom) in various applied
fields applied along the $a$-axis. $T_N $ corresponds to the
antiferromagnetic ordering temperature, and $T_1$ and $T_2$ correspond to
field-induced first- and second-order transitions respectively. The curves
are offset by 1 J/mol Ce-K$^{2}$ for each successive curve.}
\label{a115}
\end{figure}

\begin{figure}[tbp]
\caption{The cumulative phase diagrams for CeRhIn$_5$ and Ce$_2$RhIn$_8$ for
various applied fields applied along the $a$-axis. $T_N$ corresponds to the
antiferromagnetic ordering temperature, and $T_1$ and $T_2$ correspond to
field-induced first- and second-order transitions respectively. The dashed
lines are guides to the eyes. The magnetic structure in Regions I and II is
a spin density wave that is incommensurate with the lattice where Region II
has a larger magnetic moment on each Ce atom. Region III corresponds to a
spin density wave that is commensurate with the lattice.}
\label{phasediag}
\end{figure}


\begin{references}
\bibitem{firstord} The software used to measure the heat capacity uses
average values to fit the  thermal relaxation data. From the raw traces, the
transitions which we have  assigned as first order are clearly first order
in nature. Experience tells  us the averaging method underestimates the
value of the heat capacity near  the peak of the first order transition by a
factor of four or five.
\end{references}
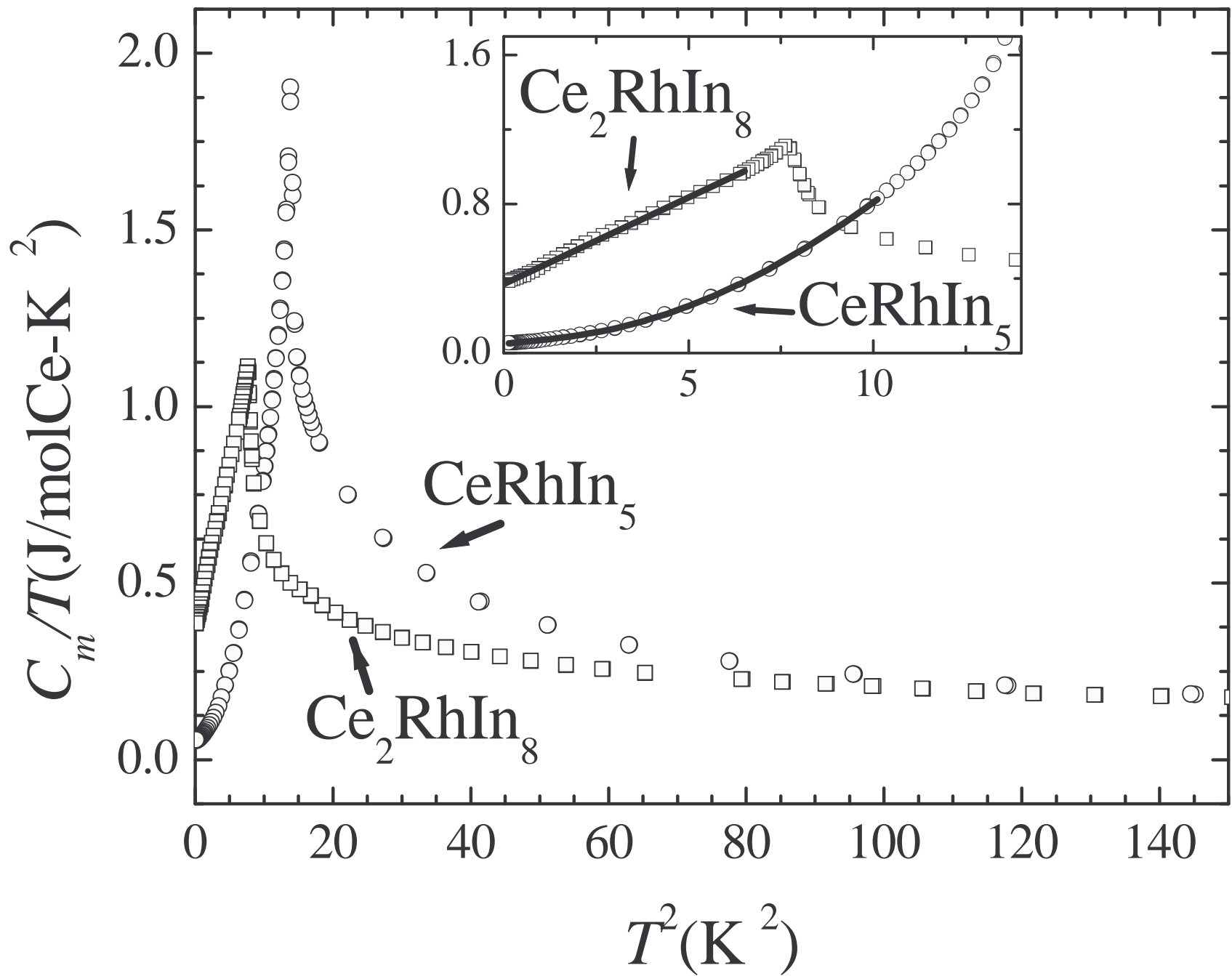
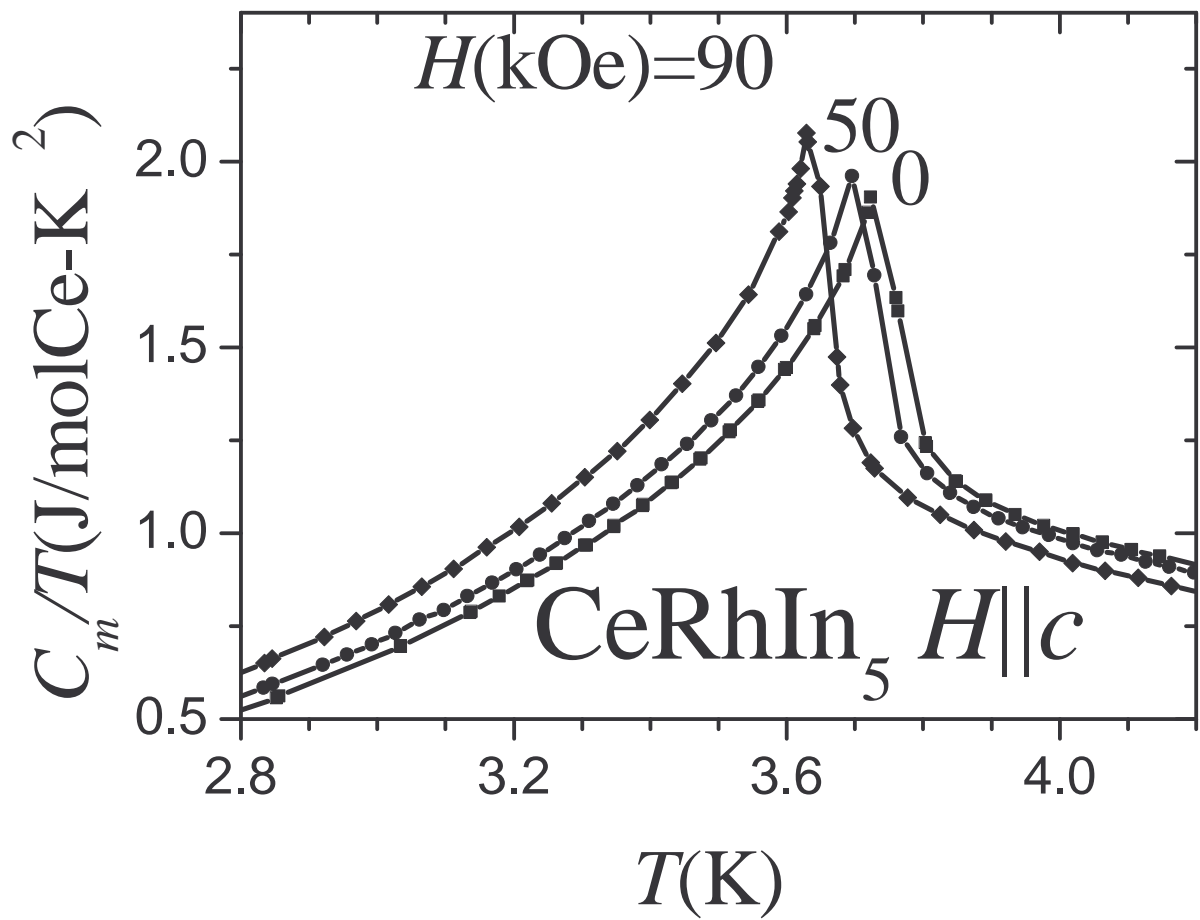
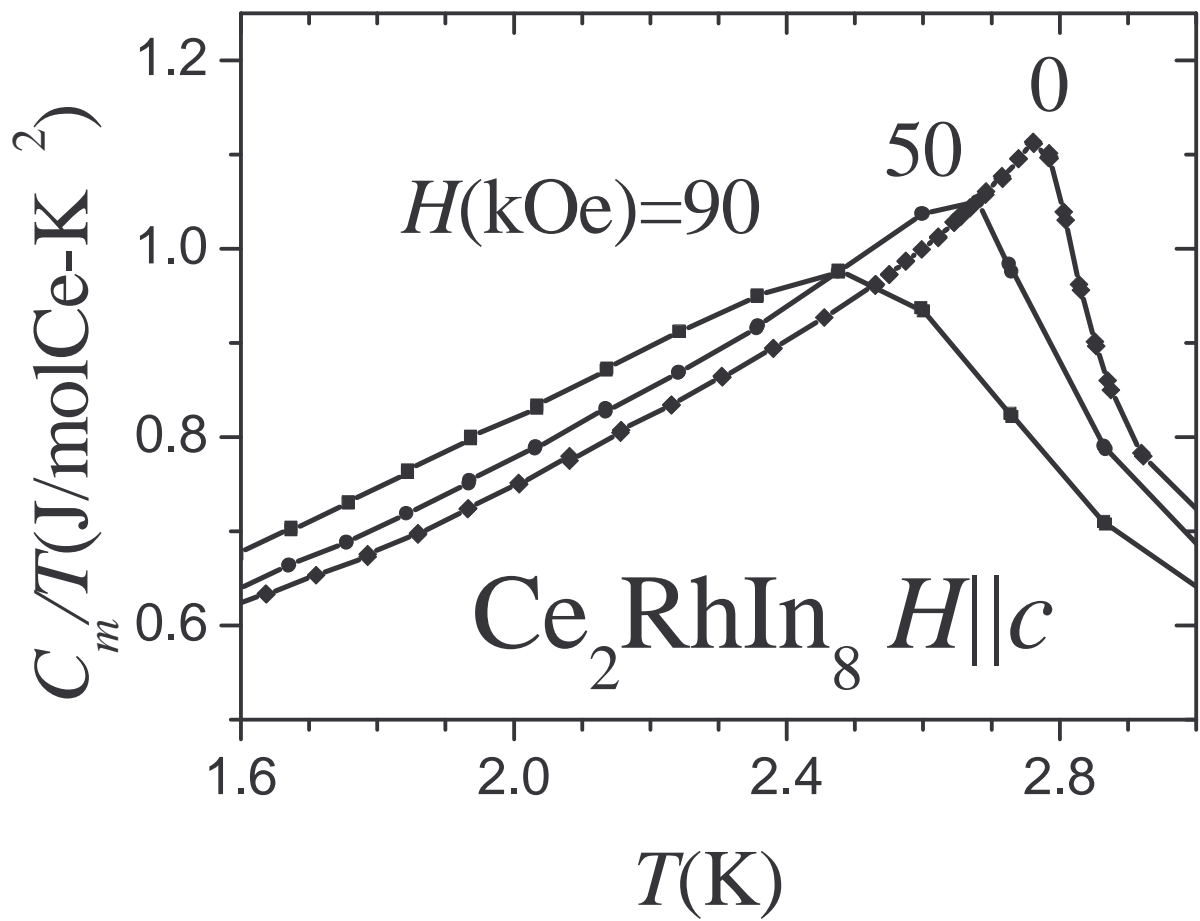
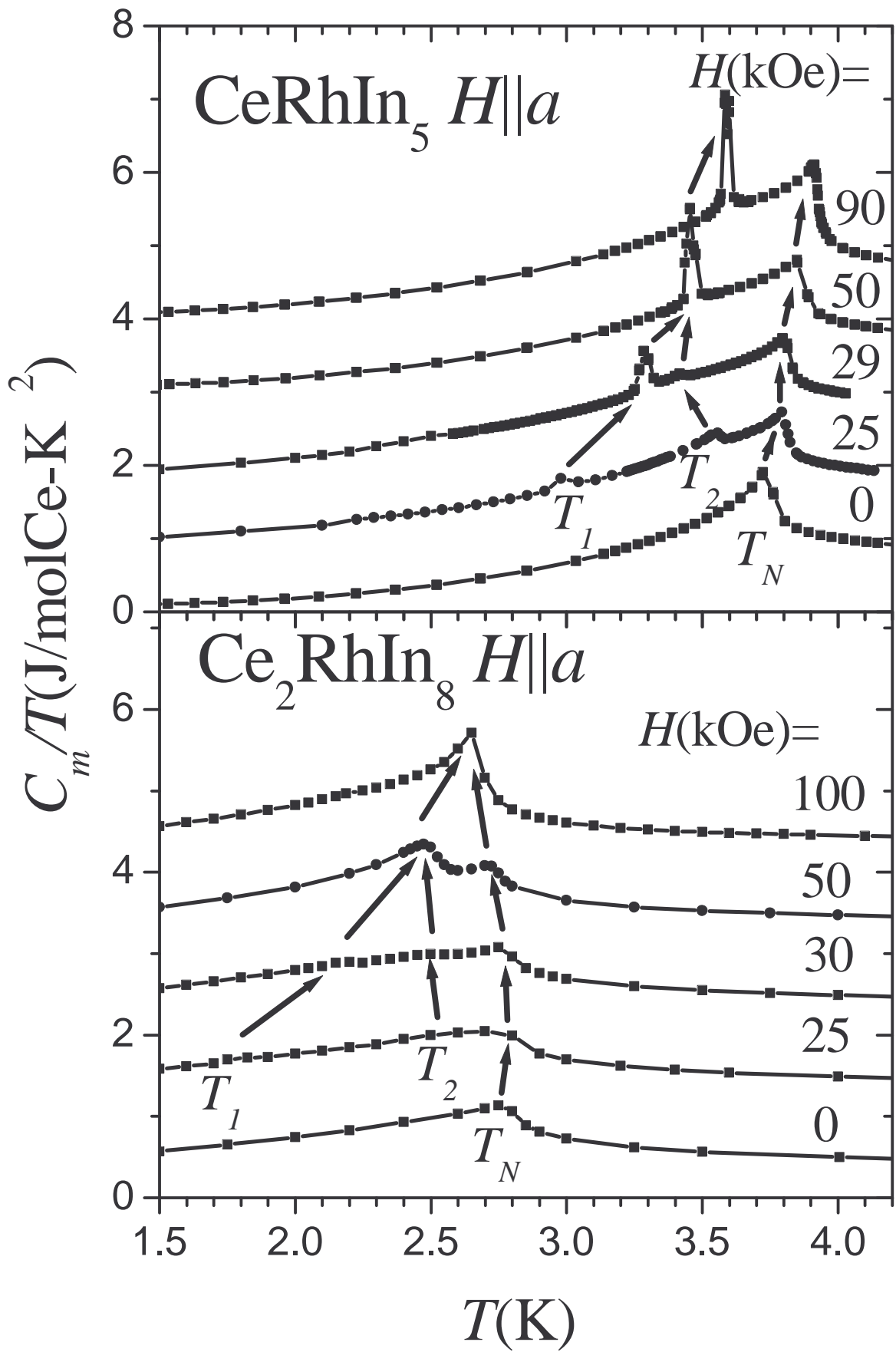
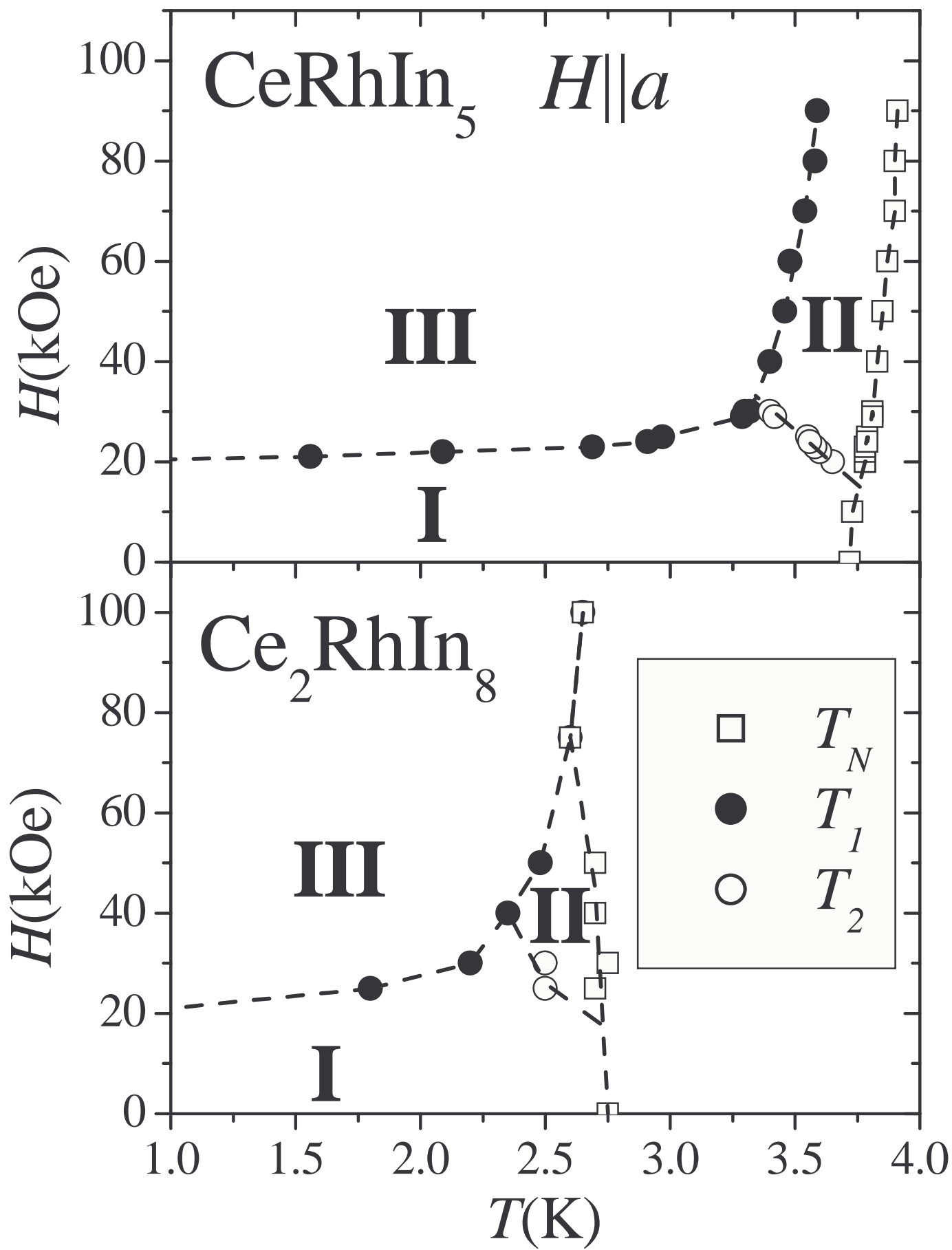
\end{document}